\documentclass[prb,aps,showpacs,twocolumn,floats,epsfig]{revtex4-1}
\usepackage{amssymb}
\usepackage{amsbsy}
\usepackage{amsmath}
\usepackage{wasysym}
\usepackage{bm}
\usepackage{epsfig}
\begin{document}

\title{Topological aspects of an exactly solvable spin chain}

\author{Abhinav Saket$^{1}$, S. R. Hassan$^{2}$ and R. Shankar$^{2}$}

\affiliation{$^{1}$Harish Chandra Research Institute, Chhatnag Road, Jhunsi, Allahabad 211 019, India}

\affiliation{$^{2}$The Institute of Mathematical Sciences, C.I.T. Campus, Chennai 600 113, India}
\date{\today}

\begin{abstract}

We analyse a spin-$1/2$ chain with two-spin interactions which shown to exactly
solvable by Lieb, Schultz and Mattis \cite{Lieb}. We show that the model can be
viewed as a generalised Kitaev model that is analytically solvable for all
defect sectors. We present an alternate proof that the defect free sector is
the ground state, which is valid for a larger parameter range.  We show that
the defect sectors have degenerate ground states corresponding to unpaired
Majorana fermion modes and that the degeneracy is topologically protected
against disorder in the spin-spin couplings. The unpaired Majorana fermions can
be manipulated by tuning the model parameters and can hence be used for
topological quantum computation. 

\end{abstract}


\maketitle

\section{Introduction}

There is much interest in the physical realisation of model systems for
topological quantum computation (TQC)\cite{tqc}. TQC is being actively explored
as a physical way to achieve fault tolerent quantum computation. In this scheme
the qubits are realised using non-abelian anyons. The braiding operations on 
them implement quantum gates robustly. One of the simplest class of 
non-abelian anyons are realised in systems with unpaired Majorana fermions 
(UMF) \cite{umf}. Kitaev presented a remarkable solvable
spin-1/2 model on a honeycomb lattice \cite{kitaevhc} which realises
non-abelian anyons made up of UMF. The model can be written in terms of
Majorana fermions in the background of $Z_2$ gauge fields and the problem
reduces to solving a theory of non-interacting Majorana fermions in the
background of static $Z_2$ gauge field configurations. Kitaev showed that the
ground state is the flux free configuration. The model has a phase which is
characterised by a topological invariant, the Chern number of the fermions, 
being equal to $\pm1$. In this phase, there are UMF trapped to each vortex 
(a $Z_2$ flux). Thus, if vortices can be created and manipulated, it is 
possible to braid the UMF.

While anyons (abelian and non-abelian) are intrinsically two dimensional
objects, Alicea {\em et. al} showed that the braiding operation of UMF
can be realised in one-dimensional wire networks \cite{alicea}. They proposed a 
realisation of such networks in semiconductor wires which can be 
engineered into a state supporting UMF at the edges.

Kitaev's honeycomb model can easily be generalised to any lattice
with coordination number three, if all the bonds can be coloured using three
colours \cite{kivelson,yang,mandal,nussinov,baskaran}. 
There are theoretical proposals to realise such systems in cold atom
systems \cite{demler} and Josephson junction quantum circuits \cite{nori}.
We had proposed and analysed such a generalised one dimensional model,
which we called the Tetrahedral model \cite{abhinav}, where we showed
that there are UMF trapped to defects and that the defects could be
created and manipulated by tuning the hamiltonian parameters. However,
this required the engineering of 3-spin interactions which is not 
easy experimentally. 

In this paper we study a simpler quantum spin-1/2 chain which involves
only 2-spin interactions and can thus, in principle, be realised in 
quantum circuits \cite{nori}. The model has been studied earlier
by Leib, Shultz and Mattis \cite{Lieb}, who showed that the model was 
exactly solvable using the Jordan-Wigner transformation. They showed
that the model has infinite conserved 2-spin operators and proved that 
the ground state was in the (suitably defined) defect free sector. In each 
defect configuration, the model reduces to an analytically solvable 
fermion hopping problem and is characterised by degenerate ground
states, the exact degeneracy depending on the defect configuration.

We solve the model exactly using  Kitaev's method \cite{kitaevhc}.  We
reproduce the previous results \cite{Lieb} and extend the validity of the
regime of the proof of the ground state being the defect free sector. We then
show that there are localised UMF zero energy modes in the sectors with
defects. The ground state degeneracy corresponds to these modes being occupied
or unoccupied. The topological nature of these zero-energy modes makes them
robust against disorder in the strengths of couplings in the Hamiltonian. By
tuning the couplings of the conserved 2-spin operators in the Hamiltonian any
defect sector can be made the ground state. Thus it is possible to manipulate
the UMF and move them along the chain. 

The rest of this paper is organised as follows. In section \ref{XYZ-Ising}, we
present the model, its conserved quantities and the Jordan-Wigner
transformation which enables us to rewrite the theory in terms of Majorana
fermions hopping in the background of static $Z_2$ gauge field configurations.
The diagonalisation of the Majorana fermion problem is described in section
\ref{diagonal}. Section \ref{exactsol} gives an exact analytic solution of the
model. In section \ref{groundstate},  we present a proof that the groundstate
lies in the defect free sector. Our proof has a larger regime of validity than
the one given previously\cite{Lieb}. It also yields an expression for fermionic
gap. Section \ref{zeromode} contains a detailed analysis of the zero modes of
the Majorana fermions. We show that the zero mode and, therefore, degeneracy in
the multiparticle spectrum is topological in origin and is robust under the
variation of the couplings in the hamiltonian. We summarise our results and
discuss them in the concluding section \ref{conclusion}.

\section{The Model}
\label{XYZ-Ising}
\begin{figure}[htb] 
\includegraphics[width=\linewidth]{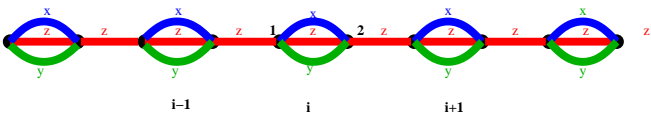}
\caption{The XYZ-Ising chain. There are two sites per unit cell. 
The $x,y$ and $z$ bonds are as indicated.}
\end{figure}
We consider the Hamiltonian,  
\begin{equation}
\label{ham}
H=\sum_{i=1}^N \left( J_x\sigma^x_{i,1}\sigma^x_{i,2}
+J_y\sigma^y_{i,1}\sigma^y_{i,2} 
+J_z\sigma^z_{i,1}\sigma^z_{i,2} 
+\sigma^z_{i,2}\sigma^z_{i+1,1}\right)
\end{equation}
with periodic boundary conditions,
\begin{eqnarray}
\label{Modelpbc}
\sigma^z_{N+1,1}=\sigma^z_{1,1}.
\end{eqnarray}

In each unit cell, there is a conserved $Z_2$ invariant, $W_i=\sigma^z_{i,1}
\sigma^z_{i,2}$. Apart from these $N$ local conserved quantities, there are
also three global conserved quantities corresponding to global $\pi$ rotations
about each of the three axes which are symmetries of the model. We denote these
by, $\Sigma^a\equiv\prod_{i=1}^N\sigma^a_{i,1}\sigma^a_{i,2}$ These are not
independent, we have $\Sigma^z=\prod_{i=1}^NW_i$ and
$\Sigma^x\Sigma^y=(-1)^N\Sigma^z$.

Thus, we see that the $J_z$ terms in the Hamiltonian (\ref{ham}) are invariants
and, hence, do not affect the eigenstates. In this work
we concentrate on $J_x=J_y\equiv\frac{J}{2}$. As we
will show, at this point the model is analytically solvable. Also, note that
this point the model has a global $U(1)$ symmetry corresponding to spin
rotations about the $z$ axis. Leib {et al} had studied the model at
the $J_z=J_x=J_y$ point and had called it the Heisenberg-Ising model. We 
follow their nomenclature and refer to the model in equation (\ref{ham})
with $J_x=J_y$ as the $XY$-Ising model.

We express the Hamiltonian in terms of Majorana fermions using the 
Jordan-Wigner transformation,
\begin{eqnarray}
\label{jwxi}
\xi_{i,1}&=&(-1)^i\sigma^z_{i,1}\mu_i,~~
\xi_{i,2}=(-1)^i\sigma^x_{i,2}\sigma^y_{i,1}\mu_i\\
\eta_{i,1}&=&\sigma^x_{i,1}\mu_i,~~
\eta_{i,2}=\sigma^z_{i,2}\sigma^y_{i,1}\mu_i
\end{eqnarray}
where $\mu_i=\prod_{i<j}\sigma^y_{j,1}\sigma^y_{j,2}$.

Then, the Hamiltonian reduces to, 
\begin{eqnarray}
\nonumber
H&=&\sum^{N-1}_{i=1}\left(i\xi_{i,2}\xi_{i+1,1}\right)+
\Sigma^x~i\xi_{N,2}\xi_{1,1}\\
\label{fham}
&&+\sum_{i=1}^N\left(J\left(\frac{1-u_i}{2}\right)i\xi_{i,1}\xi_{i,2}
+J_zu_i\right)
\end{eqnarray}
where $\hat{u}_i=-i\eta_{i,1}\eta_{i,2}$ and 
$ \prod_i^n \left(i\xi_{i,1}\xi_{i,2}\right) \prod_i^N \hat{u_i} =\Sigma^x $ 
are conserved quantities.

The Majorana operators $\xi_i$ and $\eta_i$ follow the anti-commutation relations,
\begin{eqnarray}
\{ \xi_i , \xi_j \}=2 \delta_{i,j};   
\{ \eta_i , \eta_j \}=2 \delta_{i,j}; 
\{ \xi_i , \eta_j \}=0.
\end{eqnarray}
Thus, the spin Hamiltonian gets converted into the fermionic Hamiltonian 
with periodic boundary condition when $\Sigma^x= +1$ and with 
anti-periodic boundary conditions when $\Sigma^x=-1$.

\section{Diagonalisation}
\label{diagonal}

The Hamiltonian can be diagonalised in the standard way. We write the
eigenstates as direct products of states $\vert{\cal G}\rangle$ in the $\eta$
fermion sector and states $\vert{\cal M}\rangle$ in the $\xi$ fermion sector.
We will refer to states belonging to $\eta$ fermion sector as the gauge sector
and states belonging to $\xi$ fermion sector as matter sector. We choose the
states in the gauge sector to be the simultaneous eigenstates of the $Z_2$ flux
operators, i.e $\vert{\cal G}\rangle=\vert\{u_i\}\rangle$, where

\begin{eqnarray}
\hat{u_i}\vert\{u_i\}\rangle=u_i\vert\{u_{i}\}\rangle.
\end{eqnarray}
We then have
\begin{eqnarray}
\label{sep}
H\left[\hat{u_i}\right]\vert{\cal M}\rangle
\vert\{u_i\}\rangle=
H\left[u_i\right]\vert{\cal M}\rangle
\vert\{u_i\}\rangle.
\end{eqnarray}
The single particle eigenvalue equation is given by, 
\begin{eqnarray}
\label{1defectsingle}
\frac{-i}{2}\phi_{i-1,2} + \frac{iJ(1-u_i)}{2} \phi_{i,2}=\epsilon \phi_{i,1}\nonumber\\
\frac{-iJ(1-u_i)}{2}\phi_{i,1} + \frac{i}{2} \phi_{i+1,1}=\epsilon \phi_{i,2},
\end{eqnarray}
with boundary condition
\begin{eqnarray}
\label{boundcon}
\phi_{1, 1}=\phi_{N+1, 1}.
\end{eqnarray}

We define the $u_i=-1 ~\forall ~i$ sector as the then defect free sector.  This
zero defect sector consists of two sectors corresponding to periodic or
anti-periodic boundary conditions for $\Sigma_x=+1$ and $\Sigma_x=-1$
respectively. We will prove that this is the ground state sector of the model.
If  $n$ of the $u_i$'s are equal to $+1$ then we call it the $n$-defect sector.
An $n$ defect sector reduces to solving the hopping problem on $n$ decoupled
open chains. If we can solve one defect sector (one open chain), then we can
solve all the sectors of the model because all of them reduce to decoupled
open chain Hamiltonians. 

\section{Exact Solution in all Sectors}
\label{exactsol}
In this section we put $J_z=0$ since the eigenstates are independent
of $J_z$. We first consider the zero-defect sector of a chain with
$N$ unit cells, in the thermodynamic limit $N\rightarrow\infty$. 
The Hamitonian is diagonalised in terms of the Fourier components of the 
Majorana fermion operators,
\begin{eqnarray}
\xi_i=\frac{1}{\sqrt{N}}\sum_{k}\xi_k e^{-ikR_i}
\end{eqnarray}
where $R_i=ai$, $a$ being the lattice spacing, we have supressed the 
sub-lattice index and $\xi_i$ represent
two component column vectors. The hermiticity of $\xi_i$ implies 
that $\xi_{-k}=\xi^\dagger_k$, their components satisfy the canonical 
anti-commutation relations,
\begin{equation}
\left\{\xi_{k,a},\xi^\dagger_{k',a'}\right\}=\delta_{aa'}\delta{kk'}
\end{equation}
Fermionic modes are thus defined on half the Brillouin zone. 
The momentum space Hamiltonian is,
\begin{eqnarray}
\label{momham}
H&=&\frac{1}{\pi}\int_0^{\pi/a}~dk~\xi^\dagger_k h(k)\xi_k\\
\mbox{where}
\label{spham}
h(k)&=&\left(\begin{array}{cc}
0&J-e^{ika}\\
J-e^{-ika}&0
\end{array}\right).
\end{eqnarray}
The Hamiltonian in the diagonalised form is expressed as
\begin{eqnarray}
H=\frac{1}{\pi}\int_0^{\pi/a}~dk~
\epsilon(k)\left((\chi^+_k)^\dagger{\chi}^+_{k}
-(\chi^-_k)^\dagger\chi^-_k\right)
\end{eqnarray} 
where, 
\begin{eqnarray}
\chi_k^\pm&=&\frac{1}{\sqrt 2}\left(\begin{array}{c}
1\\ \pm e^{-i\alpha_k}\end{array}\right)\\
 \tan \alpha_k&=&\frac{-\sin(ka)}{J-\cos(ka)}\\
\mbox{and~}\epsilon(k)&=&\sqrt{ J^2-2J\cos(ka)+1}.
\end{eqnarray}
Thus, the sytem is gapped for $J\ne 1$.
The values of $k$ are determined by applying the boundary 
condition (\ref{boundcon}),
\begin{eqnarray}
ka=\frac{(2m-p)\pi}{N},
\end{eqnarray} 
where $m= 1,2,3 \dots N$ and $p=0,1$ for (PBC) and anti-(PBC) respectively.

\begin{figure}[htb] 
\includegraphics[width=\linewidth]{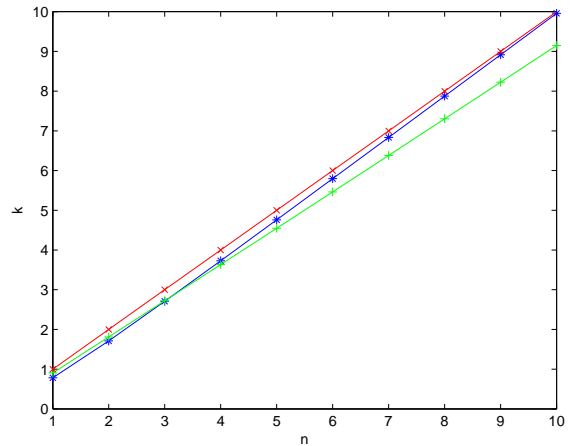}
\caption{The red line shows the plot between $k$ and $n$ without the correction $\alpha_k$. 
The green and blue line shows the plot between $k$ and $n$ for $J < 1$ and for $J > 1$ respectively.}\label{kversusn.fig}
\end{figure}

We now consider the one defect sector. The fermionic system is then an
open chain with $N$ unit cells satisfying the boundary conditions,
\begin{eqnarray}
\label{openboundcond}
\phi_{0,1}=0\nonumber\\
\mbox{and~~}\phi_{N,2}=0.
\end{eqnarray}
We use the standing waves to diagonalise this open chain single
particle Hamiltonian, 
\begin{eqnarray}
\phi_{i,a}=\frac{1}{\sqrt{2N}}\sum_{k}\phi_{k,a}e^{ikR_i} 
+ \phi_{-k,a}e^{-ikR_i}.\nonumber
\end{eqnarray}
The boundary conditions (\ref{openboundcond}) imply that
\begin{eqnarray}
\label{defectfinal}
e^{i\alpha_k}=e^{i(N+1)ka}.
\end{eqnarray}
Thus, $k$ are determined by the solution of,
\begin{eqnarray}
\label{allowed_k1}
ka=\frac{n\pi}{N+1} +\frac{\alpha_k}{N+1}.
\end{eqnarray}
We have solved equation (\ref{allowed_k1}) 
numerically for $N=10$ and plotted $k$ versus $n$ for $J>1$ and $J<1$ as 
shown in figure \ref{kversusn.fig}.  

\begin{figure}[htb] 
\includegraphics[width=\linewidth]{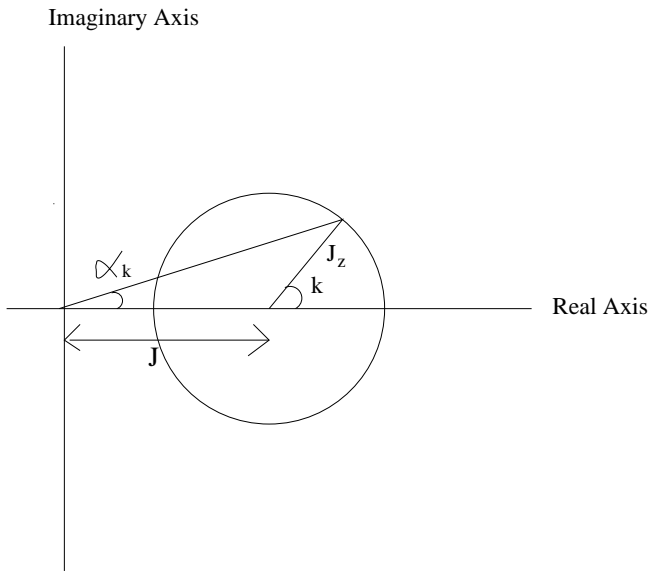}
\caption{$J- e^{ika}=\epsilon_k e^{i\alpha_k}$ in Argand plane. For $J > 1$, $\alpha(0)=0$ and $~\alpha(\pi)=0$.}\label{complexplane1}
\end{figure}

\begin{figure}[htb] 
\includegraphics[width=\linewidth]{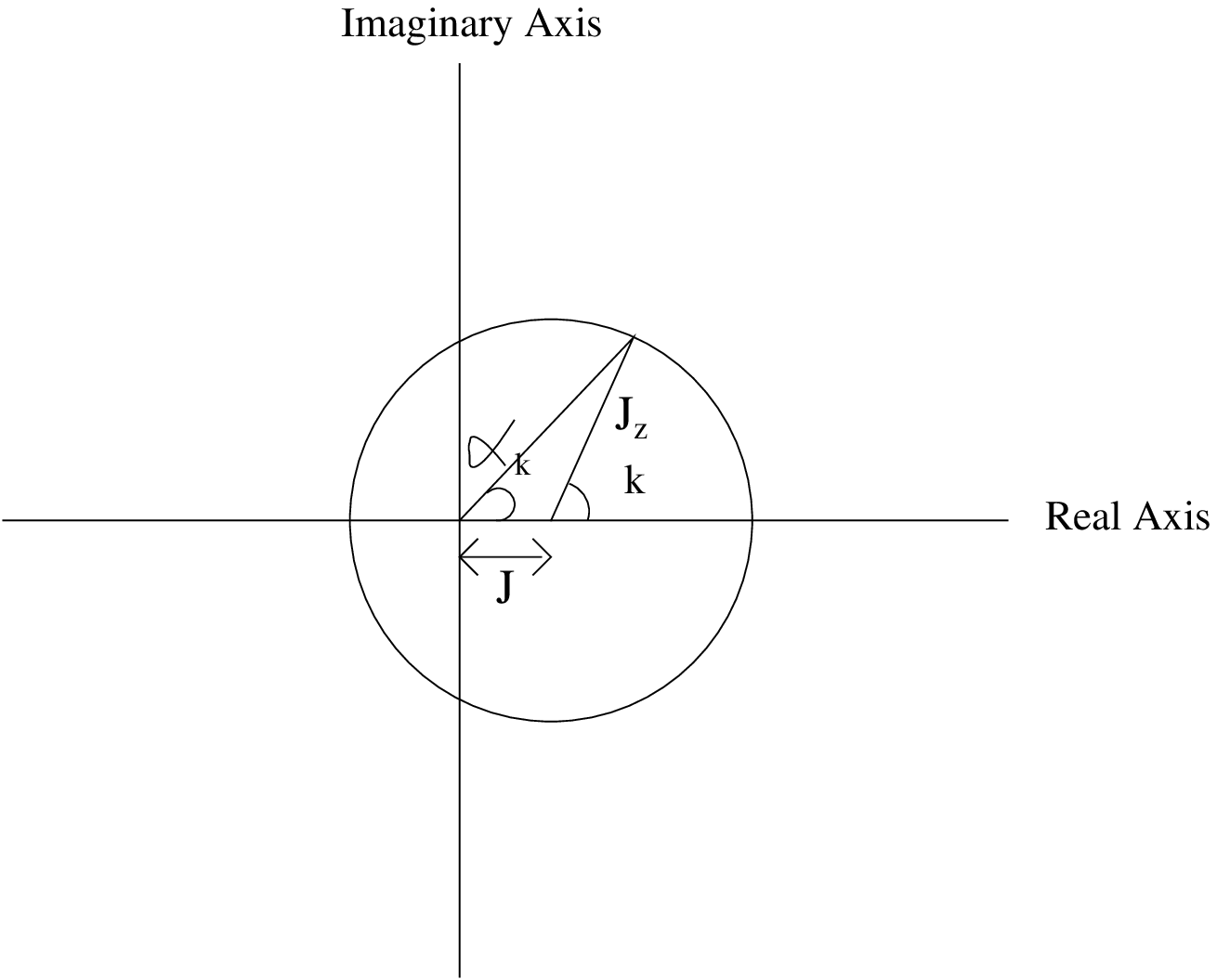}
\caption{$J- e^{ika}=\epsilon_k e^{i\alpha_k}$ in Argand plane. For $J < 1$, $\alpha(0)=0$ and $~\alpha(\pi)=\pi.$}\label{complexplane2}
\end{figure}

We see the number of allowed values of $k$ is $N$ for $J>1$ but 
is $N-1$ when $J<1$. As we will show later, the missing mode is a zero 
energy eigenfunction
of the single particle Hamiltonian with the wavefunction peaked at the edges 
of the open chain. The relation $ J-e^{ika}=\epsilon_k e^{i\alpha_k} $ has 
been shown in the Argand plane in the figure (\ref{complexplane1}) and figure
(\ref{complexplane2}). The appearance of the zero mode depends on the topology 
of parameters of Hamiltonian as shown in
figure (\ref{complexplane1}) and (\ref{complexplane2}). Zero mode appears when
the unit circle shown there encloses the origin. This argument can be 
generalised to show that if the path circles the origin $\nu$ times,
then there will be $\nu$ only $N-\nu$ standing waves.
$\nu$ can be identified with a topological invariant of the closed chain 
in terms of the Berry potential defined on the half-Brillioun zone of the 
closed chain,
\begin{equation}
A(k)=i\left(\phi_k^\dagger\partial_k\phi_k-h.c\right)
\end{equation}
where $\phi_k$ are the two-component single particle wavefunctions.
The topological invariant is the winding number of the relative phase 
of the two components of the wavefunction along the half Brillioun zone,
\begin{equation}
\nu=\frac{1}{\pi}\int_0^{\pi/a}~dk~A(k)
\end{equation}


The geometric interpretation of $\nu$ is as follows. 
Consider a general Hamiltonian of the form
\begin{eqnarray}
\nonumber
h(k)&=&\epsilon(k)\hat n(k)\\
\label{genham}
\mbox{and~}\hat n(k)&=&\cos\theta_k\tau^z
+\sin\theta\left(\cos\Omega_k\tau^x+\sin\Omega_k\tau^y\right)
\end{eqnarray}
where $\tau^a$ are the Pauli matrices. The Hamiltonian in equation
(\ref{spham}) is the case of $\theta_k=\pi/2$ and $\Omega_k=\alpha_k$.
For fully connected near neighbour chains, the off-diagonal terms 
in the single particle Hamiltonian in equation do not vanish implying
that $\theta_k\ne 0,\pi$ for any $k$. Further, if $\epsilon_k\ne 0$
the Hamitonian in equation (\ref{genham}) represents a mapping
of the half-Brillioun zone to the sphere with two points removed
(the north and south poles in our convention), which is topologically
equivalent to an annulus. Thus, $\nu$ which is the winding number of 
the map is a topological invariant for the class of gapped chains 
for which no nearest neighbour hopping amplitude vanishes.

Thus, we expect the zero energy  edge state to be robust and 
independent of disorder in the couplings, provided the system
remains gapped and fully connected by nearest neighbour
couplings. We will demonstrate this explicitly in a later section.

The solution of the open chain presented above can be used to
to solve $n$ defect sector, for any $n$ as these sectors
consist of $n$ decoupled open chains.

\section{Ground state and Gap}
\label{groundstate}

In this section, we will prove that zero defect sector is the ground state
sector for $J_z=0$. In order to prove the statement we first prove that the
ground state energy of zero defect sector is less than that of one defect
sector for large $N$. We then prove that ground state energy of one defect
sector is less than or equal to that of the $n$-defect sector.  We then turn on
$J_z$ and compute regime of the stability of the system to defect formation. We
finally compare our results with those of Lieb {\em et. al.}\cite{Lieb} and
point out precisely how our proof extends the regime of validity of their
result.

{\bf First step:} We outline the proof that the ground state energy of 
zero defect sector $GSE_{zds}$ is less that of the one defect sector,
$GSE_{ods}$. The details are given in Appendix \ref{proof}. The ground
state energies of the two sectors are,
\begin{eqnarray}
\label{gsezds}
GSE_{zds}&=&-\sum_{k=1}^N \epsilon \left(\frac{2k\pi}{N}\right)\\
\label{gseods}
GSE_{ods}&=&-\sum_{k=1}^N \epsilon \left(\frac{k\pi}{N+1}
+\frac{\alpha(k\pi/N+1)}{N+1}\right)
\end{eqnarray}
In the thermodynamic limit of $N\rightarrow\infty$, we convert the 
summations into integrals using the Euler-Maclaurin formula to get,
\begin{eqnarray}
GSE_{zds}&=&-\frac{N}{\pi}\int_0^{\pi} \epsilon(k)dk\\
\nonumber
GSE_{ods}&=&-\frac{N}{\pi}\int_{0}^{\pi} \epsilon(k)dk + J(1-\delta(J))\\
\delta(J)&=&\frac{1}{\pi}\int_{0}^{\pi}dk
\frac{\cos(k)-J}{\sqrt{J^2- 2J\cos(k)+ 1}} 
\end{eqnarray}
We then show in Appendix that $\delta(J)<1$ for all finite $J$, thus proving that the ground state energy of zero defect sector is less than that of the one defect sector.

{\bf Second step:} Now, we show that the ground state energy of one defect 
sector is less than that of two defect sector and so on. To show this, 
we split up Hamiltonian into three parts,
\begin{eqnarray}
H_{L_1+L_2}=H_{L_1}+H_{L_2}+ H_{12} \label{gse2ds2}
\end{eqnarray}
where H is the Hamiltonian for one defect sector, $H_{L_1}$ and $H_{L_2}$ are Hamiltonians 
of lengths $L_1, L_2$. $H_{12}$ is the link bond between $H_{L_1}$ and $H_{L_2}$.

From variational principle, 
\begin{eqnarray}
\epsilon_G \left( L_1+L_2 \right) \leq \langle \psi|H_{L_1}+H_{L_2}+ H_{12}|\psi\rangle\label{gse2ds4}
\end{eqnarray}
where $\epsilon_G\left(L_1+L_2\right)$ is the ground state energy of Hamiltonian for one defect sector.

Using the trial wave function $|\psi\rangle = \vert \psi_{L_1}\rangle \vert \psi_{L_2}\rangle$, we get
\begin{eqnarray}
\epsilon_G \left( L_1+L_2 \right) &\leq& \epsilon_G \left( L_1\right) + \epsilon_G \left( L_2 \right)\nonumber\\ 
&+& \langle\psi_{L_1}\vert \langle \psi_{L_2}\vert H_{12} \vert \psi_{L_1}\rangle \vert \psi_{L_2}\rangle 
\end{eqnarray}
Let us consider the link Hamiltonian $H_{12}=i\xi_n\xi_{n+1}$  where $\xi_n$ belongs to the Hamiltonian $H_{L_1}$ and $\xi_{n+1}$ belongs to the Hamiltonian $H_{L_2}$. Then,
\begin{eqnarray}
\langle\psi_{L_1}\vert \langle \psi_{L_2}\vert H_{12} \vert \psi_{L_1}\rangle \vert \psi_{L_2}\rangle=i\langle\psi_{L_1}\vert \xi_n \vert \psi_{L_1}\rangle\langle\psi_{L_2}\vert \xi_{n+1} \vert \psi_{L_2}\rangle\nonumber.
\end{eqnarray}
Since the expectation value of single $\xi_n$ operator in the ground state is zero. Therefore,
\begin{eqnarray}
\epsilon_G\left(L_1+L_2\right) \leq \epsilon_G \left( L_1\right) + \epsilon_G \left( L_2 \right) 
\end{eqnarray}

Thus, we have proved that the ground state energy of one defect sector is less
than the ground state energy of two defect sector. Similarly, we can prove for
two defect sector and so on. Therefore, zero defect sector is the ground state
sector at $J_z=0$. 

We now turn on $J_z$ and examine the region of stability of the zero defect 
sector under defect production. The difference of the ground state energies
of the zero and one defect sector for non-zero $J_z$ is
\begin{equation}
GSE_{zds}-GSE_{ods}= J(1-\delta(J))+2J_z
\end{equation}
Thus, the system will become unstable towards defect production when
\begin{equation}
J_z<-\frac{J}{2}\left(1-\delta(J)\right).
\end{equation}
This extends the result of Lieb {\em et. al.} since their proof is not
valid for negative values of $J_z$.

\subsection{Nature of low energy excitations}

\begin{figure}[htb] 
\includegraphics[width=\linewidth]{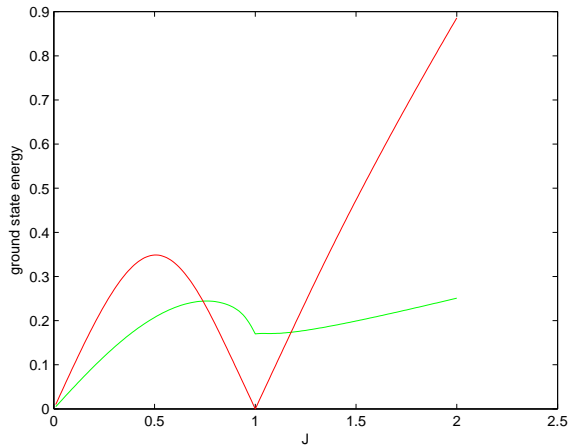}
\caption{Low energy excitation in XY-Ising Model. The ground state energy of one defect sector is shown in green and first excitation of zero defect sector is shown in red. While plotting all the energies, ground state of zero defect sector has been taken as reference.}\label{lowenergyexcitation.fig}
\end{figure}

For $J_z>-\frac{J}{2}(1-\delta(J))$, the zero defect is the ground state sector.
The first excited state can be either the first excitation in the 
zero defect sector or the ground state of the one defect sector. 
The excitation gap in the zero defect sector follows from the single
particle spectrum and is given by
\begin{eqnarray}
\Delta=|(1-J)|.
\end{eqnarray} 
Therefore, model is gapless for $J=1$. Numerically, we have shown that for $J_z=0$ first excited state of the model in the region $0<J\le 0.75$ and $J\ge 1.17$ is the ground state of the one defect sector . The excited state of the zero defect sector is the first excited state in the region $0.74\le J\le 1.17$. These results have been shown in figure (\ref{lowenergyexcitation.fig}). We also see that the model has a finite gap for all non-zero values of $J\ne 1,0$.  

\section{Zero modes}
\label{zeromode}
We now analyse the zero mode of the Majorana fermions. When $\epsilon=0$, 
equations (\ref{1defectsingle}) decouple and become
\begin{eqnarray}
-\frac{J(1-u_i)}{2}\phi_{i,1}+ \phi_{i+1,1}&=&0\nonumber\\
\label{zm1}
\mbox{and~~}-\phi_{i-1,2} + \frac{J(1-u_i)}{2}\phi_{i,2}&=&0.
\end{eqnarray}
These recursion relations can formally be solved,
\begin{eqnarray}
\phi_{i,1}&=&\prod_{j=1}^i\left(\frac{J(1-u_j)}{2}\right)\phi_{1,1}\nonumber\\
\label{phi1sol}
\mbox{and~~}\phi_{i,2}&=&\prod_{j=i}^N\left(\frac{2}{J(1-u_j)}\right)\phi_{N,2}.
\end{eqnarray}
There are two degenerate and independent solutions for every set of values of 
the parameters and every flux configuration. In zero defect sector, imposing 
boundary condition $\phi_{1,1}=\phi_{N+1,1}$, we find that the zero modes exist 
for zero defect sector only at the gapless point $J=1$. In one defect sector, 
which is an open chain with boundary condition $\phi_{0,1}= \phi_{N,2}= 0$, 
$\phi_{i,1}$ and $\phi_{i,2}$ can be expressed as,
\begin{eqnarray}
\phi_{i,1}&=&{\left(J\right)}^i\phi_{0,1}\nonumber\\
\label{phi3sol}
\mbox{and~~}\phi_{i,2}&=&{\left(\frac{1}{J}\right)}^{i-1}\phi_{N,2}.
\end{eqnarray}
We can see that if $J>1$, then $\phi_{i,2}$ becomes zero at the boundary for 
infinite open chain. Now, if we take $\phi_{i,1}=0 \forall i$, then the 
boundary condition at the other end $\phi_{0,1}=0$ is also satisfied for $u_1=-1$.
Therefore, the only one solution for zero mode exist in an infinite chain 
for $J>1$. The same is true for the case $J<1$. In both cases, we get only 
one non-zero solution for zero mode.

Thus, the zero mode occurs when the toplogical invariant is non-zero.
It can also be seen from the recursive solution presented above that the
zero modes exist even if the coupling, $J$ varies with $i$. Thus the
zero mode topologically protected against random, spatial variations in
the exchange coupling.

\section{Conclusion}
\label{conclusion}
To summarize, we have analysed an exactly solvable spin-1/2 chain which was
earlier studied by Lieb, Shultz and Mattis \cite{Lieb}. We have shown that
it is a generalised Kitaev chain. Using the methods of Kitaev \cite{kitaevhc}
we can reproduce the earlier results. The formalism also helps us extend
the earlier proof to negative vlaues of $J_z$ and compute the value of
$J_z$ when the defect free state becomes unstable to defect production.
In addition we have explicitly displayed the topological nature of the zero 
energy modes in the presence of defects. We have identified the topological 
invariant that determines the existence of zero energy edge modes in the 
system. Thus the degeneracy is protected against random variations in the 
exchange coupling.

In conclusion, our results show that zero energy Majorana modes can be 
created and manipulated in this model exactly in the same way shown in 
previous work\cite{abhinav}. The system has a gap for all non-zero 
values of $J\ne1$. Thus the UMF can used as a topological qubit. 
The importance of this model is that it has only two-spin couplings
which can, in principle, be realised experimentally\cite{nori}.

\section*{Acknowledgments}
We thank Deepak Dhar for a very useful discussion.

\appendix

\section{Details of proof}
\label{proof}

The ground state energy $GSE_{zds}$ of zero defect sector is expressed as  
\begin{eqnarray}
GSE_{zds}=-\sum_{k=1}^N \epsilon\left(\frac{2k\pi}{N}\right),\label{gsezds1}
\end{eqnarray}
where $\epsilon(k)=\sqrt{ J^2- 2J\cos(ka)+1}$.

To evaluate the summation series we convert it into integral using Euler-Maclaurin formula,
\begin{eqnarray}
\sum_{x=1}^n f(x)&=&\int_0^n f(x)dx +B_1[f(0)-f(n)]\nonumber\\
&+&\sum_{k=1}^p\frac{B_{2k}}{(2k)!}\left(f^{(2k-1)}(n)-f^{(2k-1)}(0)\right) 
\end{eqnarray}
where $B_1 = -1/2, B_2 = 1/6, B_3 = 0, B_4 = -1/30, B_5 = 0, B_6 = 1/42...$ are the Bernoulli numbers.

The expression for ground state energy for zero defect sector equation (\ref{gsezds1}) in thermodynamic limit $N \rightarrow \infty$ becomes,
\begin{eqnarray}
GSE_{zds}=-\frac{N}{\pi}\int_0^{\pi} \epsilon(y)dy. 
\end{eqnarray}
where we have used the identity $\epsilon(0)=\epsilon(2\pi)=J+J_z$. 

The ground state energy $GSE_{ods}$ of one defect sector is, 
\begin{eqnarray}
GSE_{ods}=-\sum_{k=1}^N \epsilon \left(\frac{k\pi}{N+1}+\frac{\alpha(k\pi/N+1)}{N+1}\right)\label{gseods}
\end{eqnarray}

To evaluate the summation series, we convert it into integral using Euler-Maclaurin formula and neglect last terms containing (N+1) in the denominator in limit $N \rightarrow \infty$,
\begin{eqnarray}
GSE_{ods}&=&-\int_0^N \epsilon\left(\frac{k\pi}{N+1} + \frac{\alpha(\frac{k\pi}{N+1})}{N+1}\right)dk\nonumber\\
&+&\frac{1}{2}\epsilon\left(\frac{\alpha(0)}{N+1} \right)\nonumber\\ 
&-&\frac{1}{2}\left[\epsilon\left(\frac{N\pi}{N+1} + \frac{\alpha(\frac{N\pi}{N+1})}{N+1}\right) \right].\label{gseods2}
\end{eqnarray}
In the thermodynamic limit $N \rightarrow \infty$, above equation (\ref{gseods2}) becomes, 
\begin{eqnarray}
GSE_{ods}&=&-\int_0^N \epsilon\left(\frac{k\pi}{N+1}+\frac{\alpha(k\pi/N+1)}{N+1}\right)dk\nonumber\\
&+&\frac{1}{2}\left[\epsilon(0)-\epsilon(\pi)\right].
\end{eqnarray}
Let us substitute $y=\frac{k\pi}{N+1}$, and above expression for ground state energy for one defect sector becomes
\begin{eqnarray}
GSE_{ods}&=&-\frac{N+1}{\pi}\int_{0}^{\frac{N\pi}{N+1}} \epsilon\left(y +\frac{\alpha(y)}{N+1}\right)dy\nonumber\\
&+&\frac{1}{2}\left[\epsilon(0)-\epsilon(\pi)\right].
\end{eqnarray}
Using Taylor expansion and neglecting terms containing $\frac{1}{N+1}$, 
\begin{eqnarray}
GSE_{ods}&=&-\frac{N+1}{\pi}\int_{0}^{\frac{N\pi}{N+1}} \epsilon(y)dy\nonumber\\ 
&-&\frac{1}{\pi}\int_{0}^{\frac{N\pi}{N+1}}\left(\alpha(y)\epsilon'(y)\right)dy.
\end{eqnarray}

Using integration by parts in the second term and then taking the thermodynamic limit, we get
\begin{eqnarray}
GSE_{ods}=&-&\frac{N}{\pi}\int_{0}^{\pi} \epsilon(y)dy\nonumber\\ 
&-& \frac{1}{\pi}\int_{0}^{\pi}dy\frac{J\cos(y)-J^2}{\sqrt{J^2- 2J\cos(y)+ 1}}\nonumber\\
&+& \frac{1}{2}\left[ \epsilon(0)+\epsilon(\pi) \right]\nonumber\\
&+& \frac{1}{\pi} \left[ \alpha(\pi)\epsilon(\pi)+\alpha(0)\epsilon(0)\right].
\end{eqnarray}
From figures (\ref{complexplane1}, \ref{complexplane2}), 
\begin{eqnarray}
\epsilon(0)&=&J+1\nonumber\\
\epsilon(\pi)&=&J-1~~when~~J > 1 \nonumber\\
\epsilon(\pi)&=&1-J~~when~~J < 1 \nonumber\\
\alpha(0)&=&0                    \nonumber\\
\alpha(\pi)&=&0~~when~~J > 1   \nonumber\\
\alpha(\pi)&=&\pi~~when~~J < 1. 
\end{eqnarray}
Substituting all these values, we show that for both cases $J>1$ and $J<1$ ground state energy of one defect sector is expressed as,
\begin{eqnarray}
GSE_{ods}&=&GSE_{zds} + J(1-\delta(J))\nonumber\\ 
\mbox{where~}\delta(J)&=&\frac{1}{\pi}\int_{0}^{\pi}dy\frac{\cos(y)-J}{\sqrt{J^2 - 2J\cos(y)+1}}\label{finstate}.
\end{eqnarray}
Since in the expression for $\delta(J)$ the integrand is always less than 1, we can write
\begin{eqnarray}
\delta(J)&<&\frac{1}{\pi}\int_{0}^{\pi}dy=1.
\end{eqnarray}
Therefore, from equation (\ref{finstate}), we can say that the ground state energy of zero defect sector is less than ground state energy of one defect sector.

\end{document}